\documentclass[aps,reprint,superscriptaddress]{revtex4-1}

\usepackage{graphicx} 
\usepackage{amsmath,bm} 

\begin{document}
\title{Constrained quantum annealing of graph coloring} 

\author{Kazue Kudo}
\email[]{kudo@is.ocha.ac.jp}
\affiliation{Department of Computer Science, Ochanomizu University,  
Tokyo 112-8610, Japan}

\date{\today}

\begin{abstract}
We investigate a quantum annealing approach based on real-time quantum dynamics for graph coloring.
In this approach, a driving Hamiltonian is chosen so that constraints are naturally satisfied without penalty terms, and the dimension of the Hilbert space is considerably reduced.
The total Hamiltonian, which consists of driving and problem Hamiltonians, resembles a disordered quantum spin chain.
The ground state of the problem Hamiltonian for graph coloring is degenerate.
This degeneracy is advantageous and is characteristic of this approach.
Real-time quantum simulations in a small system demonstrate interesting results and provide some insight into quantum annealing.
\end{abstract}

\maketitle

\section{Introduction}
\label{sec:intro}

Quantum annealing (QA)~\cite{kadowaki1998,brooke1999,das2008}, 
which is also known as adiabatic quantum computation~\cite{albash2018},
is a quantum-mechanical approach for optimization problems~\cite{farhi2001,martonak2004,kurihara2009,altshuler2010,titiloye2011,titiloye2012}.
QA has attracted considerable interest from a wide variety of fields since experimental results by the first commercial quantum annealers were reported~\cite{johnson2011}.
In a QA framework, the Hamiltonian consists of two parts:
a problem Hamiltonian $H_{\rm p}$ and a driving Hamiltonian $H_{\rm d}$.
The initial state is taken as the ground state of $H_{\rm d}$, 
which is supposed to be unique and easily prepared.
The total Hamiltonian is changed from $H_{\rm d}$ to $H_{\rm p}$,
\begin{equation}
 H(s)=sH_{\rm p} + (1-s)H_{\rm d},
\end{equation}
where $s(t)$ is a time-dependent parameter and $0\le s \le 1$.
The QA process starts from $s=0$ at $t=0$ and ends with $s=1$ at time $t=T$.
If the process is adiabatic, the solution of an optimization problem is obtained as the ground state of $H_{\rm p}$.
The time required for an adiabatic process is proportional to a polynomial of the inverse of the energy gap between the ground and first-excited states.
A number of attempts to avoid exponentially small energy gaps have been proposed for accelerating an adiabatic process~\cite{albash2018,nishimori2017,hormozi2017,susa2018}. 

QA can be simulated by either a quantum Monte Carlo (QMC) method or by solving the time-dependent Schr\"{o}dinger equation.
The QMC method (the path-integral Monte Carlo method, in particular) is often used as a QA approach. 
Although relatively large systems can be simulated using this method, the simulation dynamics do not represent true real-time quantum evolution.  
In contrast, the time-dependent Schr\"{o}dinger equation provides true quantum dynamics.
However, solving the Schr\"{o}dinger equation requires significantly more computational resources than the QMC method. 
QA based on real-time Schr\"{o}dinger evolution has another difficulty in that practical optimization problems often include constraints.
The standard approach to imposing constraints is to include penalty terms in the problem Hamiltonian~\cite{gatian2012,bian2013,gaitan2014,lucas2014,rieffel2015}.
In real quantum devices, constraints are accommodated by using minor embedding techniques~\cite{choi2008,vinci2015}.
These approaches are often disadvantageous for QA performance.
Recently, another approach called constrained quantum annealing (CQA) was proposed~\cite{hen2016a,hen2016b}, and an extension of the approach called Quantum Alternating Operator Ansatz was also proposed~\cite{hadfield2017a,hadfield2017b}.
In the CQA approach, the driver Hamiltonian is chosen so that it commutes with the constraint operators but not with the problem Hamiltonian.
This approach, which utilizes the symmetry of the system, naturally restricts the Hilbert space to a subspace with a considerably small dimension.

In this paper, we focus on graph coloring, which is one of the optimization problems in which
CQA is effective.
For example, in a standard QA approach using the transverse-field Ising model, 
real-time quantum simulation requires $2^{qN}$ dimensions to color a graph with $N$ nodes with $q$ colors.
However, in a CQA approach, the dimension of the Hilbert space reduces to only $q^N$.
The dimension reduction is essential for the simulation of real-time quantum dynamics since it leads to a reduction of memory usage and computation time.
The number of avoided crossings (small energy gaps) is small in a small Hilbert space, since the number of energy levels is small.
Therefore, the dimension reduction is expected to result in better performance.
Although the size of the system to which CQA is applicable is still smaller than that of the QMC method, real-time Schr\"{o}dinger evolution has an advantage over the QMC method in solving a graph coloring problem.
When a graph is colorable with $q$ colors, there are at least $q!$ solutions.
In other words, the ground state of the problem Hamiltonian is degenerate. 
Although this type of degeneracy is resolved in the formulation proposed in Refs.~\cite{titiloye2011,titiloye2012}, a standard QMC method is ineffective in such a highly-degenerate case~\cite{kurihara2009}.
By contrast, in real-time quantum dynamics, the degenerate ground state is favorable, since populations of several low-energy states merge at the end of an annealing process.

The rest of the paper is organized as follows.
In Sec.~\ref{sec:model},
models, which consist of problem and driving Hamiltonians, are presented, and methods of numerical simulations are outlined in detail.
Two types of driving Hamiltonians are also introduced.
One consists of chains with nearest-neighbor (NN) coupling, and the other consists of fully-connected (FC) chains.  
Results are presented on the dependence of the annealing time in Sec.~\ref{sec:annealing}.
The residual energy and success probability, which measure the performance of QA, are examined.  
These two quantities are compared for NN and FC cases using two different annealing methods: linear annealing and exponential annealing.
In Sec.~\ref{sec:gap}, the time evolution in the linear-annealing case is analyzed.
The time dependence of the ground-state population and population distributions demonstrate interesting behavior of real-time quantum dynamics.
Conclusions are highlighted in Sec.~\ref{sec:conc}.

\section{Models and methods}
\label{sec:model}

Graph coloring is coloring the nodes of a graph such that no pair of nodes connected by an edge has the same color.
The classical Hamiltonian for coloring a graph $G=(V,E)$ with $q$ available colors is given by
\begin{equation}
 H_{\rm cl} =\sum_{(ij)\in E}\sum_{a=1}^q
\frac{S_{i,a}+1}{2}\frac{S_{j,a}+1}{2},
\label{eq:H_cl}
\end{equation}
where $i$ and $j$ represent nodes $V=\{1,\ldots,N\}$, 
and $(ij)\in E$ denotes the edge which connect the pair of nodes $i,j\in V$.
If node $i$ is colored $a$, the Ising spin variable is $S_{i,a}=1$; otherwise, $S_{i,a}=-1$.
Since one node has only one color, the constraint of this problem is expressed as
\begin{equation}
 \sum_{a=1}^q S_{i,a} = 1\cdot 1+(-1)\cdot(q-1) = 2-q,
\label{eq:const}
\end{equation}
for $i=1,\ldots,N$.
Next, consider the quantum version of the Hamiltonian~\eqref{eq:H_cl}, namely, the problem Hamiltonian. We define the problem Hamiltonian as
\begin{equation}
 H_{\rm p}=J\sum_{(ij)\in E}\sum_a \sigma^z_{i,a}\sigma^z_{j,a},
\label{eq:H_p}
\end{equation}
where $\sigma^z_{i,a}$ denotes the Pauli matrix, and the first and second indices are for node and color, respectively.
We take $J=1$, which has a unit of energy.
Here, constant terms and a factor $1/4$ which arise from Eq.~\eqref{eq:H_cl} are eliminated for convenience.
When a given graph with $m$ edges is colorable with $q$ colors, the ground-state energy of $H_{\rm p}$ is $E_0=m(q-4)$.
If we choose an extended $XY$ model as a driving Hamiltonian, we have
\begin{equation}
 H_{\rm d}=-J\sum_i\sum_{a,b}
\left(\sigma^x_{i,a}\sigma^x_{i,b}+\sigma^y_{i,a}\sigma^y_{i,b}\right),
\label{eq:H_d}
\end{equation}
then the total Hamiltonian becomes a kind of $XXZ$ model.
Since $\langle\sum_{a=1}^q\sigma^z_{i,a}\rangle$ is conserved in this model, 
the constraint corresponding to Eq.~\eqref{eq:const} is naturally satisfied if the initial state is prepared accordingly.

\begin{figure}[tb]
 \includegraphics[width=8cm,clip]{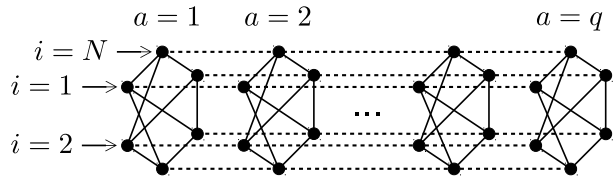}%
\caption{\label{fig:chain}
Schematic of the model which consists of Eqs.~\eqref{eq:H_p} and \eqref{eq:H_d}. 
The solid lines represent the edges of a graph, namely, the interaction in the problem Hamiltonian.
The broken lines correspond to the driving Hamiltonian.
}
\end{figure}

In a special case, the total Hamiltonian corresponds to, in a sense, an ensemble of one-dimensional tight-binding chains.
If the summation over $a$ and $b$ in Eq.~\eqref{eq:H_d} is restricted to $b=a+1$, $H_d$ is merely the summation of $N$ independent $XY$ chains.
Now that each chain has only one up spin, it is equivalent to a tight-binding chain.
As depicted in Fig.~\ref{fig:chain}, each site of a tight-binding chain (represented as a broken line) interacts with the corresponding site of other chains through the problem Hamiltonian (represented by solid lines).
If we focus on one of the chains, it resembles a tight-binding chain with random local potentials.
We refer to this driving Hamiltonian as a nearest-neighbor (NN) type.
Although this type is sufficient to realize CQA, additional connections between nodes in each chain may enhance the efficiency of transition.
For comparison, we also consider a fully-connected (FC) type in which the summation over $a$ and $b$ in Eq.~\eqref{eq:H_d} is taken for all combinations of $a<b$.

In order to investigate the performance of QA, numerical simulations are performed in the cases where one or more solutions exist.
In Sec.~\ref{sec:annealing}, we consider four-coloring ($q=4$) of regular random graphs of connectivity $c=3$, which is far below the colorable/uncolorable transition threshold.
Graphs of connectivity below the threshold are expected to be colorable in the large-$N$ limit.
The colorable/uncolorable threshold for regular random graphs is $c_s=10$ (for $q=4$) ~\cite{krzakala2004,zdeborova2007}. 

In the following sections, we consider only the subspace that satisfies a constraint $\langle\sum_{a=1}^q\sigma^z_{i,a}\rangle =2-q$ and refer to the lowest-energy state in the subspace as the ground state.
The initial condition is given by the lowest-energy state of $H_{\rm d}$ in the subspace.
Note that the initial state is not the global ground state.
If the initial state is prepared in the whole Hilbert space of the system (i.e., without restriction to the subspace), one can add an additional Zeeman term with an appropriate magnetic field to achieve the desired state, as mentioned in Ref.~\cite{hen2016b}. 
Time evolution is calculated by solving the time-dependent Schr\"{o}dinger equation by the 4-th Runge-Kutta method.
Two different annealing schedules are examined: 
linear annealing, $s(t)=t/\tau_{\rm li}$, and exponential annealing, $s(t)=1-\exp(-t/\tau_{\rm ex})$.
The final time for linear annealing is $T=\tau_{\rm li}$, which leads to $s(T)=1$.
For exponential annealing, however, $s=1$ at $T\to\infty$.
To reach sufficiently close to $s=1$, we take $T=15\tau_{\rm ex}$, and then $1-s(T)<10^{-6}$.
Time is measured in units of $\hbar/J$.

\section{Dependence on annealing time}
\label{sec:annealing}

\begin{figure}[tb]
 \includegraphics[width=8cm,clip]{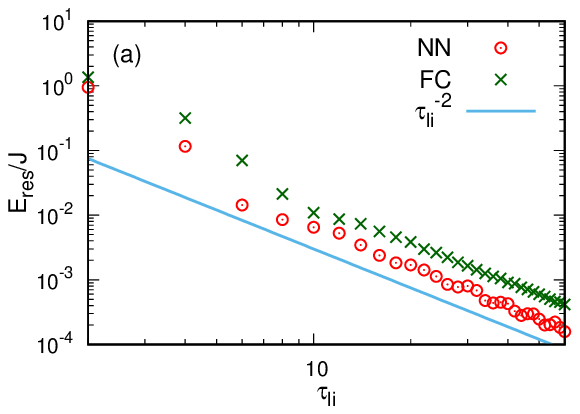}
 \includegraphics[width=8cm,clip]{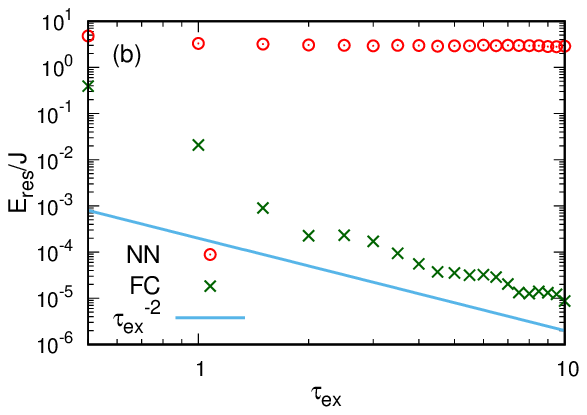}%
\caption{\label{fig:tc} (Color online)
Annealing-time dependence of residual energy $E_{\rm res}$
for (a) linear annealing and (b) exponential annealing.
Data for two different driving Hamiltonians are compared:
nearest-neighbor (NN) and fully-connected (FC) types.
If the annealing process is adiabatic, the residual energy is expected to obey the power law with respect to the annealing time: 
$E_{\rm res}/J\propto \tau_{\rm li}^{-2}$, $\tau_{\rm ex}^{-2}$.
}
\end{figure}

In this section, we consider four-coloring of regular random graphs with several solutions.
Numerical simulations are performed for 100 realizations of random graphs with $N=6$ nodes and connectivity $c=3$.
The performance of QA is often demonstrated by the residual energy, which is the energy difference between the expected value at the final time and the ground-state energy of the problem Hamiltonian.
The residual energy is defined by
\begin{equation}
 E_{\rm res}=\langle\Psi_{\rm f}| H_{\rm p} |\Psi_{\rm f}\rangle -E_0,
\label{eq:E_res}
\end{equation}
where $|\Psi_{\rm f}\rangle$ is the final state, and $E_0$ is the ground-state energy of $H_{\rm p}$. 
The residual energy averaged over 100 realizations is plotted as a function of the annealing time in Fig.~\ref{fig:tc}.
For an adiabatic QA process, the residual energy is expected to obey a power law with respect to the annealing time $\tau$, specifically, 
$E_{\rm res}/J\propto \tau^{-2}$~\cite{suzuki2005}.
In Fig.~\ref{fig:tc}(a), the case of linear annealing, 
$E_{\rm res}/J\propto\tau_{\rm li}^{-2}$ in a large-$\tau_{\rm li}$ regime for both the NN and FC types of driving Hamiltonians.
However, in Fig.~\ref{fig:tc}(b), the case of exponential annealing, 
the power-law behavior appears only for the FC type.
For the NN type, $E_{\rm res}/J$ maintains a large value even in a large-$\tau_{\rm ex}$ regime.

\begin{figure}[tb]
 \includegraphics[width=8cm,clip]{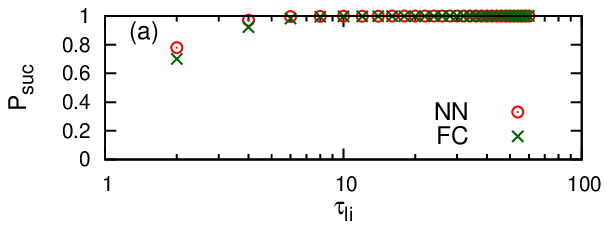}
 \includegraphics[width=8cm,clip]{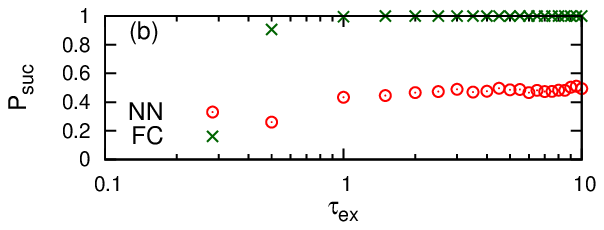}%
\caption{\label{fig:tp} (Color online)
Annealing-time dependence of the success probability $P_{\rm suc}$
for (a) linear annealing and (b) exponential annealing.
Data for nearest-neighbor (NN) and fully-connected (FC) types of driving Hamiltonians are compared.
}
\end{figure}

Another practical measure is success probability.
This is the probability that a selected answer at the end of QA is one of the solutions. 
The final state can be written as the superposition of state bases: 
$|\Psi_{\rm f}\rangle=\sum_i c_i|i\rangle$, 
where $H_{\rm p}$ is diagonalized with $|i\rangle$'s.
Each state basis $|i\rangle$ denotes a combination of $z$-components of spins, namely, a color combination of nodes.
When $\langle i|H_{\rm p}|i\rangle=E_0$, $|c_i|^2$ represents the probability that the $i$th state (namely, color combination) is a solution.
Therefore, the success probability is defined as
\begin{equation}
 P_{\rm suc}=\sum_i|c_i|^2\delta(\langle i|H_{\rm p}|i\rangle,E_0),
\label{eq:P_suc}
\end{equation}
where $\delta(\cdot,\cdot)$ denotes the Kronecker delta.
The success probability averaged over 100 realizations as a function of the annealing time is shown in Fig.~\ref{fig:tp}.
The success probability for linear annealing is quite high in a large-$\tau_{\rm li}$ regime, as shown in Fig.~\ref{fig:tp}(a). 
This result is consistent with the result of Fig.~\ref{fig:tc}(a).
However, in Fig.~\ref{fig:tp}(b), the case of exponential annealing, the success probability for the NN type is much lower than that of the FC type.
This low success probability is demonstrated by the large residual energy in Fig.~\ref{fig:tc}(b) for the NN type. 

\begin{figure}[tb]
 \includegraphics[width=8cm,clip]{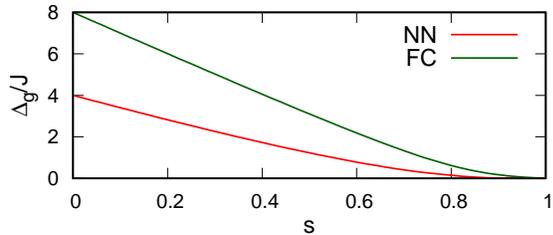}%
\caption{\label{fig:gap} (Color online)
Energy gap $\Delta_{\rm g}(s)/J$ between the ground and first-excited states of the Hamiltonian of an example. 
The graph corresponding to the problem Hamiltonian is given by Fig.~\ref{fig:graph}(a).
Data for the nearest-neighbor (NN) and the fully-connected (FC) types driving Hamiltonians are compared.
}
\end{figure}

According to the adiabatic theorem of quantum mechanics, a quantum state evolves adiabatically if the evolution time $T$ satisfies the following condition~\cite{das2008,albash2018}:
\begin{equation}
 T\gg 
\frac{\displaystyle\max_{0\le s\le 1}
\bigg|\left\langle\phi_1(s)\left|\frac{dH(s)}{ds}\right|\phi_0(s)\right\rangle\bigg|}
{\displaystyle\min_{0\le s\le 1}\Delta_g^2(s)},
\quad
s=t/T,
\label{eq:ad}
\end{equation}
where $|\phi_0(s)\rangle$ and $|\phi_1(s)\rangle$ are the instantaneous ground state and first-excited state, respectively, and $\Delta_g(s)$ is the instantaneous energy gap between the two states.
In the system considered here, the ground state is degenerate at $s=1$.
Whenever $0\le s <1$, the gap of the FC type is larger than that of the NN type, as shown in Fig.~\ref{fig:gap}.
We naturally expect that the performance of QA for the FC type should be better than the NN type.
This expectation appears to be correct for exponential annealing, as shown in Figs.~\ref{fig:tc}(b) and \ref{fig:tp}(b).
However, Figs.~\ref{fig:tc}(a) and \ref{fig:tp}(a) show opposite results.
This unexpected behavior in the linear-annealing case is analyzed in the next section.

\section{Time evolution in a linear-annealing case}
\label{sec:gap}

\begin{figure}[tb]
 \includegraphics[width=6cm,clip]{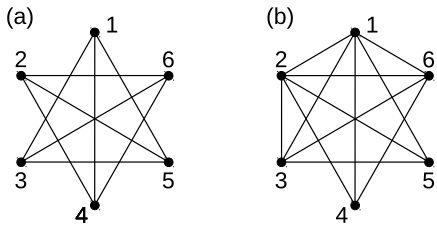}%
\caption{\label{fig:graph}
Examples of graphs that correspond to the problem Hamiltonian.
(a) Regular random graph with connectivity $c=3$.  
(b) Graph with only $4!$ degenerate solutions in a four-coloring problem.
}
\end{figure}

In this section, we focus on linear annealing and investigate what causes the unexpected result of the previous section.
We take Fig.~\ref{fig:graph}(a) as an example of a regular random graph with connectivity $c=3$.
In the case of four-coloring, the ground state of $H_{\rm p}$ is highly-degenerate for this graph.
In contrast, for Fig.~\ref{fig:graph}(b), the degree of degeneracy is only $4!$.
Note that the ground state of the total Hamiltonian is nondegenerate for $0\le s<1$ and degenerate only at $s=1$.

\begin{figure}[tb]
 \includegraphics[width=8cm,clip]{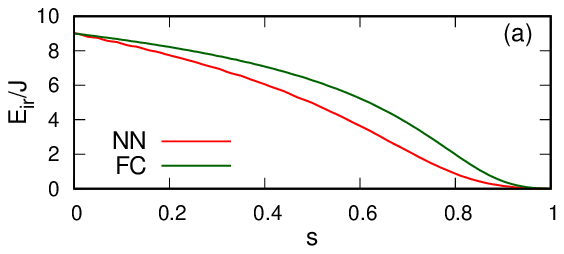}
 \includegraphics[width=8cm,clip]{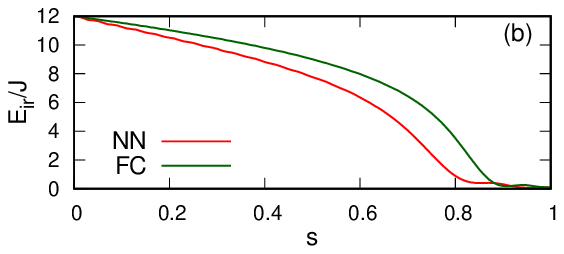}%
\caption{\label{fig:t_cost} (Color online)
Time evolution of the instantaneous residual energy, which is measured in units of $J$, for NN-type and FC-type driving Hamiltonians.
Annealing time is $\tau_{\rm li}=20$.
The problem Hamiltonians for (a) and (b) correspond to Figs.~\ref{fig:graph}(a) and \ref{fig:graph}(b), respectively. 
}
\end{figure}

Time evolution of the instantaneous residual energy is defined by
\begin{equation}
 E_{\rm ir}(s)=\langle\Psi(s)|H_{\rm p}|\Psi(s)\rangle -E_0,
\end{equation}
where $|\Psi(s)\rangle$ is the wavefunction at $s$, and is illustrated in Fig.~\ref{fig:t_cost}.
Here, the annealing time is taken as $\tau_{\rm li}=20$.
In Fig.~\ref{fig:t_cost}(a), $E_{\rm ir}$ is almost zero at $s=1$, which is consistent with the result in Fig.~\ref{fig:tc}(a).
The residual energy at $s=1$ in Fig.~\ref{fig:t_cost}(b) is slightly larger than that shown in Fig.~\ref{fig:t_cost}(a).
This is because the graph of Fig.~\ref{fig:graph}(b) has more edges, and thus, coloring is more difficult in Fig.~\ref{fig:graph}(b) than Fig.~\ref{fig:graph}(a).
Note here that a faster decay in $E_{\rm ir}$ does not mean better performance of QA because $E_{\rm ir}$ depends on the total Hamiltonian, which is different for the NN and FC types.

\begin{figure}[tb]
 \includegraphics[width=8cm,clip]{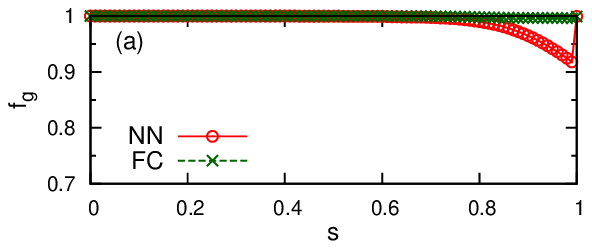}
 \includegraphics[width=8cm,clip]{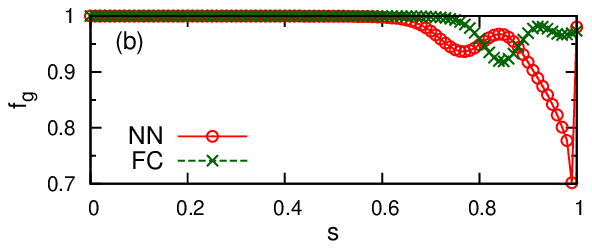}%
\caption{\label{fig:t_pg} (Color online)
Time evolution of the population of the instantaneous ground state, 
$f_g(s)=\langle\phi_0(s)|\phi_0(s)\rangle$,
for NN-type and FC-type driving Hamiltonians.
Annealing time is $\tau_{\rm li}=20$.
The problem Hamiltonians for (a) and (b) correspond to Figs.~\ref{fig:graph}(a) and \ref{fig:graph}(b), respectively. 
}
\end{figure}

A peculiar behavior is observed during the time evolution of the ground-state population, as shown in Fig.~\ref{fig:t_pg}, which is defined by
\begin{equation}
 f_g(s)=\langle\phi_0(s)|\phi_0(s)\rangle,
\end{equation}
where $|\phi_0(s)\rangle$ is the instantaneous ground state of the total Hamiltonian.
When a QA process is adiabatic, $f_g=1$ all the time.
In Fig.~\ref{fig:t_pg}(a), $f_g$ maintains a value of almost $1$ for the FC type.
In contrast, for the NN type, $f_g$ apparently decays from $s\simeq 0.8$, and jumps back to $1$ at $s=1$.
A similar behavior appears in Fig.~\ref{fig:t_pg}(b):
$f_g$ for the NN type starts to decay earlier than the FC type and jumps back to approximately $1$ at $s=1$.
This peculiar behavior is caused by the degeneracy of the ground state at $s=1$.
Since the energy gap for the NN type is smaller than that of the FC type, it is natural that the ground-state population becomes smaller for the NN type than for the FC type.
During an annealing process, the population can transfer to low excited states.
Some of the lowest-excited states merge to the ground state at the end.
This causes a jump of the ground-state population at $s=1$.

\begin{figure}[tb]
 \includegraphics[width=4.2cm,clip]{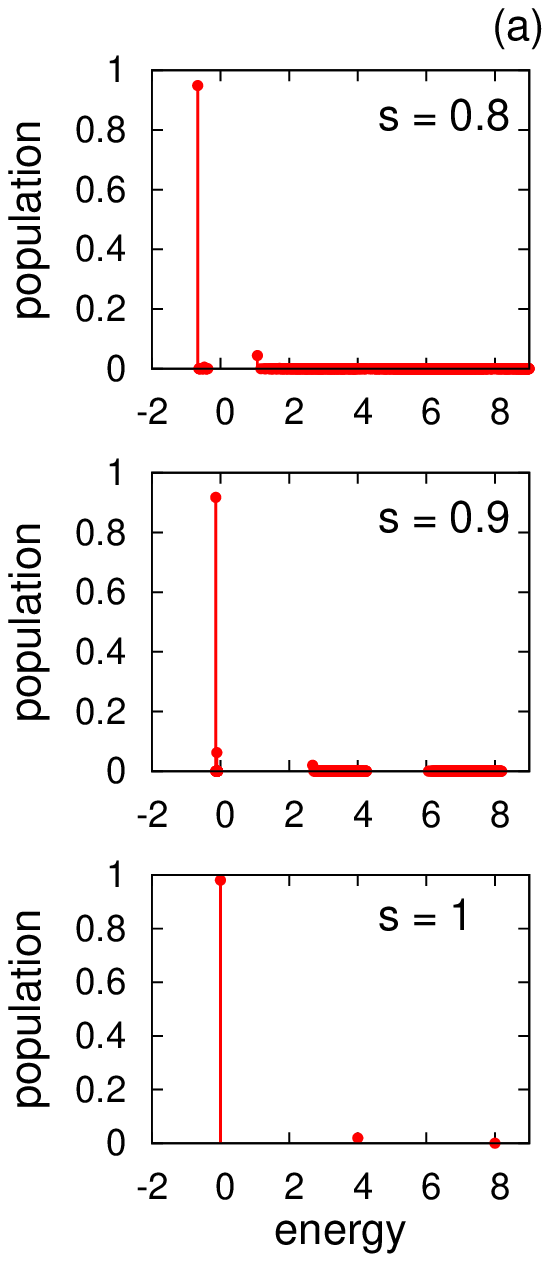}
 \includegraphics[width=4.2cm,clip]{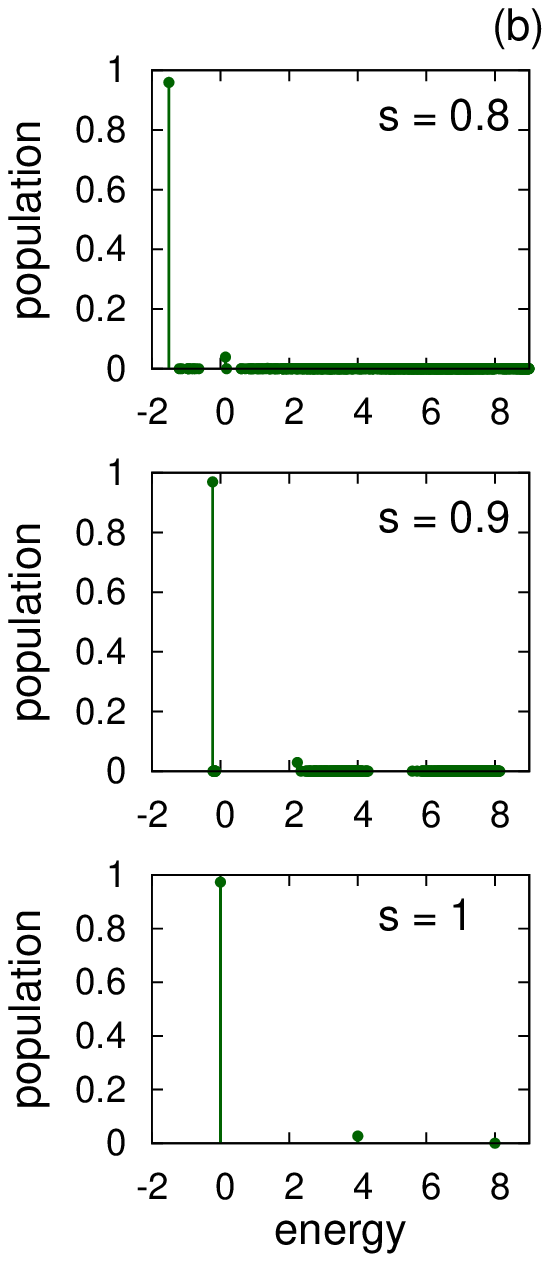}%
\caption{\label{fig:pops} (Color online)
Population distributions in a low-energy range for (a) NN-type and (b) FC-type driving Hamiltonians.
The problem Hamiltonian corresponds to Fig.~\ref{fig:graph}(b).
Annealing time is $\tau_{\rm li}=20$.
Energy is measured in units of $J$.
}
\end{figure}

The population distributions in Fig.~\ref{fig:pops} demonstrate how the ground-state and low-excited-state populations change in the last stage.
The data in Fig.~\ref{fig:pops} correspond to Fig.~\ref{fig:t_pg}(b), which shows a larger difference between the NN and FC types than Fig.~\ref{fig:t_pg}(a).
The population distributions at $s=0.8$ for the NN type [Fig.~\ref{fig:pops}(a)] and the FC type [Fig.~\ref{fig:pops}(b)] have similar properties.
The ground-state population is the largest, and a low-excited state has a small fraction.
At $s=0.9$, the ground-state population is smaller for the NN type than the FC type. However, an excited state that is very close to the ground state has a small but recognizable fraction for the NN type.
The small fraction merges to the ground-state population at $s=1$, and the population distributions for the NN and FC types appear to be almost the same.
This is the scenario of the abrupt jump of $f_g$ at $s=1$ in Fig.~\ref{fig:t_pg}.

At this point we still have an unanswered question: 
why is the residual energy smaller for the NN type relative to the FC type in the linear annealing case?
Figure~\ref{fig:pops} gives some insights.
Although the energy gap between the ground and first-excited states is smaller for the NN type, the gaps between the energy bands are smaller for the FC type.
This implies that the population spreads to the excited states which eventually merge to an excited state of $H_{\rm p}$ more easily for the FC type than the NN type.
A smaller population in higher excited states results in a smaller residual energy for the NN type than the FC type.

\section{Conclusions}
\label{sec:conc}

Real-time quantum dynamics using a CQA approach has some advantages in regard to solving a graph coloring problem.
In particular, the reduction in the dimension of the Hilbert space is a remarkable advantage of CQA.
Small dimensions are favorable for searching for solutions and are of practical convenience for simulations. 
When a graph is colorable with $q$ colors, there are at least $q!$ solutions.
In other words, the ground state is degenerate.
The degeneracy is advantageous in real-time quantum evolution, since populations of low-excited states merge with the ground state at the end of QA.
This implies that we are likely to obtain a solution, even if an annealing process is less adiabatic.
The model used in this paper, which is a kind of $XXZ$ model, resembles disordered tight-binding or fully-connected chains.
This analogy indicates a possibility of analysis from the viewpoint of disordered low-dimensional quantum systems.
The interesting results for linear annealing suggest that real-time quantum dynamics in many-body systems may reveal unexpected or overlooked properties of quantum annealing.

\begin{acknowledgments}
This work is partially supported by JSPS KAKENHI Grant Number JP18K11333, and the research grant from the Inamori Foundation.
\end{acknowledgments}


\end{document}